\documentclass[%        Class options:
groupedaddress,%
aps,%                   American Physical Society
prd,%                   Physical Review D
showpacs,%              Displays PACS after abstract
preprint,%              Preprint layout
tightenlines,%          Single spaced lines
nofootinbib,%           Does not treat footnotes as references
floatfix]%              Fixes float errors
{revtex4}%              REVTEX 4 Package used
\usepackage{graphicx,%  Default Latex 2eps package for embedding figures
longtable,%             Useful for long table
}

\usepackage{color} %This is essential for placing graphics onto
\usepackage{multirow} % To use multirow command in tables
\usepackage[T1]{fontenc}

\newcommand{\eps}  {\varepsilon}
\newcommand{\epp}  {\varepsilon'}
\renewcommand{\baselinestretch}{1.5}
\newcommand{\AddrAHEP}{%
 AHEP Group, Institut de F\'{\i}sica Corpuscular --
 C.S.I.C./Universitat de Val{\`e}ncia \\
 Edificio Institutos de Paterna, Apt. 22085, E--46071 Valencia, Spain}
\newcommand{\AddrHamb}{%
II. Institut f\"ur theoretische Physik, Universit\"at Hamburg, Luruper Chausee 149, 22761 Hamburg, Germany}

\begin{document}
\title{Probing non-standard neutrino-electron interactions\\
with solar and reactor neutrinos}

\author{A. Bola\~nos$^1$
\email{azucena@fis.cinvestav.mx}}
\author{O. G. Miranda$^1$
\email{Omar.Miranda@fis.cinvestav.mx}}
\author{A. Palazzo$^2$
\email{valle@ific.uv.es}}
\author{M. A. T\'ortola$^3$
\email{mariam.tortola@desy.de}}
\author{J. W. F. Valle$^2$
\email{valle@ific.uv.es}}

\address{
$^1$ Departamento de F\'{\i}sica, Centro de Investigaci{\'o}n
y de Estudios Avanzados del IPN,\\
Apt. Postal 14-740 07000 M\'exico, D F, M\'exico\\
$^2$\AddrAHEP\\
$^3$\AddrHamb}

\begin{abstract}

 Most neutrino mass extensions of the standard electroweak model
 entail non-standard interactions which, in the low energy limit, can
 be parametrized in term of effective four-fermion operators
 $\nu_\alpha \nu_\beta \bar f f $.  Typically of sub-weak strength,
 $\epsilon_{\alpha \beta} G_F$, these are characterized by
 dimensionless coupling parameters, $\epsilon_{\alpha \beta}$, which
 may be relatively sizeable in a wide class of schemes. Here we focus
 on non-universal (NU) flavor conserving couplings ($\alpha = \beta$)
 with electrons ($f = e$) and analyse their impact on the
 phenomenology of solar neutrinos.  We consistently take into account
 their effect both at the level of propagation where they modify the
 standard MSW behavior, and at the level of detection, where they
 affect the cross section of neutrino elastic scattering on electrons.
 We find limits which are comparable to other existing
 model-independent constraints.
 \end{abstract}

\pacs{14.60.Pq, 13.15.+g}

% 14.60.Pq : Neutrino mass and mixing 
% 13.15.+g : Neutrino interactions
\maketitle

\section{Introduction}

Solar neutrino oscillations dominated by matter effects
\cite{wolfenstein:1978ue,mikheev:1985gs} are currently well
established by solar neutrino experiments
\cite{cleveland:1998nv,Fukuda:1996sz,abdurashitov:2002nt,hampel:1998xg,Altmann:2005ix,gallex-nu08,Gavrin-talk-nu08,fukuda:2001nj,fukuda:2001nk,fukuda:2002pe,Smy:2003jf,ahmad:2001an,ahmad:2002jz,ahmad:2002ka,Ahmed:2003kj,Aharmim:2005gt,Aharmim:2008kc,
Arpesella:2007xf, Arpesella:2008mt, Collaboration:2008mr} and have
been confirmed by the long-baseline KamLAND reactor
experiment~\cite{eguchi:2002dm,araki:2004mb,:2008ee}. The combination
between solar and KamLAND determines a unique solution in the
mass-mixing parameter space, the so-called Large Mixing Angle (LMA)
solution, see
e.g.~\cite{Fogli:2005cq,Maltoni:2004ei,Schwetz:2008er,Fogli:2008ig}. This
solution has been shown to be quite robust against possible
uncertainties in solar physics, such as magnetic fields in the
radiative zone, that could give rise to noise
fluctuations~\cite{Schaefer:1987fr,Krastev:1991hc,Loreti:1994ry,nunokawa:1996qu,Burgess:2003fj,burgess:2002we,Burgess:2003su,Fogli:2007tx},
as well as in the convective
zone~\cite{miranda:2000bi,Miranda:2003yh}, that could induce
spin-flavor neutrino
conversions~\cite{schechter:1981hw,akhmedov:1988uk}.
The KamLAND data play a crucial role in establishing that non-standard
effects can play only a subleading role~\cite{pakvasa:2003zv}, their
amplitude being effectively constrained.

Altogether, the high precision and robustness of the current data
render solar and reactor neutrinos a unique probe of possible physics
beyond the Standard
Model~\cite{pakvasa:2003zv,Berezhiani:2001rt,Davidson:2003ha,Guzzo:2004ue,Friedland:2004pp,Barranco:2005ps,Barranco:2007ej},
complementing information from atmospheric and accelerator
neutrinos~\cite{fornengo:2001pm,Friedland:2004ah}. Moreover
non-standard interactions provide an important window of opportunity
for current or upcoming long-baseline neutrino oscillation
experiments, and have been extensively considered in this
framework~\cite{huber:2001de,huber:2002bi,huber:2001zw,Ribeiro:2007ud,Kopp:2007rz,Kopp:2008ds,EstebanPretel:2008qi}.

It is worth stressing that, while constrained by the solar and KamLAND
data, non-standard interactions (NSI) provide an exception
to robustness of the
neutrino oscillation 
interpretation~\cite{Friedland:2004pp,Guzzo:2004ue} and they might even
shift the solution to the so--called dark side region of the neutrino
parameter space~\cite{Miranda:2004nb}.  
Indeed, with oscillations still being the underlying
mechanism, an additional degenerate oscillation solution in neutrino
oscillation parameters can appear for sufficiently intense non-standard
interactions

Neutrino NSI constitute an unavoidable feature of gauge models of
neutrino mass, for example models of the generic seesaw
type~\cite{schechter:1980gr} where neutrino masses arise from the
admixture of isodoublet and isosinglet neutral leptons.  In general,
the lepton mixing matrix for charged currents is described by a
matrix, $K$, and the corresponding neutral weak interactions are
described by a non-trivial matrix~\cite{schechter:1980gr} $K^\dagger
K$.  In particular, in the simplest type-I seesaw
schemes~\cite{Minkowski:1977sc,gell-mann:1980vs,yanagida:1979,mohapatra:1980ia},
the smallness of neutrino mass implies that, barring fine-tuning, the
magnitude of neutrino NSI and its effects are expected to be
negligible. However this need not be always the case. For example, by
a suitable symmetry one may prevent the appearance of type-I seesaw
mass contributions, hence allowing for the new neutral heavy leptons
to lie at a mass scale accessible to accelerator experiments and,
simultaneously, potentially produce sizeable NSI strengths. For
example, this may happen in some specially designed triplet (type-II)
seesaw models~\cite{schechter:1980gr,schechter:1982cv}, as shown in
Ref.~\cite{Gu:2008yj}.

Alternatively, one may extend the lepton sector of the $SU(2) \otimes
U(1)$ theory by adding a set of {\sl two} 2-component isosinglet
neutral fermions in each
generation~\cite{mohapatra:1986bd,Gonzalez-Garcia:1989rw}.  This
scheme is sometimes called ``inverse seesaw'' an provides an elegant
way to generate small neutrino masses without a super-heavy scale.
This automatically allows for a sizeable magnitude of neutrino NSI
strengths, unconstrained by the smallness of neutrino
masses~\footnote{It also provides an explicit example for flavour and
CP violation completely detached from the smallness of neutrino
masses ~\cite{bernabeu:1987gr,branco:1989bn,rius:1990gk}.}.
The NSI which are engendered in this case will necessarily affect
neutrino propagation properties in matter, an effect that may be
resonant in certain
cases~\cite{valle:1987gv,nunokawa:1996tg,EstebanPretel:2007yu}.
They may also be large enough as to produce effects in the laboratory.

Another possible way to induce neutrino NSI is in the context of
low-energy supersymmetry without R-parity
conservation~\cite{Hall:1984id,Ross:1984yg,Ellis:1984gi,santamaria:1987uq}
both of the
bilinear~\cite{Diaz:1997xc,Hirsch:2000ef,abada:2001zh,Diaz:2003as} and
trilinear type~\cite{barger:1989rk}.
The smallness of neutrino masses may also follow from its radiative
nature~\cite{babu:1988ki,zee:1980ai}, allowing for possibly sizeable
NSI strengths~\footnote{For an alternative recent discussion of
possible NSI strengths in a similar context see
Refs.~\cite{Gavela:2008ra, Antusch:2008tz}}.

In general one may consider a general class of non-standard
interactions described via the effective four fermion Lagrangian,
\begin{equation}
-{\cal L}^{eff}_{\rm NSI} =
\varepsilon_{\alpha \beta}^{fP}{2\sqrt2 G_F} (\bar{\nu}_\alpha \gamma_\rho L \nu_\beta)
( \bar {f} \gamma^\rho P f ) \,,
\label{eq:efflag}
\end{equation}
where $G_F$ is the Fermi constant and $\varepsilon_{\alpha
 \beta}^{fP}$ parametrize the strength of the NSI. The chiral projectors
 $P$ denote $\{R,L=(1\pm\gamma^5)/2\}$, while $\alpha$ and $\beta$ denote the
three neutrino flavors, $e$, $\mu$ and $\tau$ and $f$ 
is a first generation SM fermion ($e, u$ or $d$). 

For example, the existence of effective neutral current interactions
contributing to the neutrino scattering off $d$-quarks in matter,
provides new flavor-conserving as well as flavor-changing terms for
the matter potentials of neutrinos.  Such NSI are directly relevant
for solar~\cite{Miranda:2004nb,Guzzo:2004ue,Bergmann:1998rg} and
atmospheric neutrino
propagation~\cite{gonzalez-garcia:1998hj,Friedland:2004ah,fornengo:2001pm}.

In general, the presence of NSI affects the solar neutrino
phenomenology inducing profound modifications both in matter
propagation~\cite{valle:1987gv,guzzo:1991hi,roulet:1991sm} as well as
in the detection process~\cite{Berezhiani:2001rt}.
Although various works have investigated the effects of NSI at the
level of propagation inside the
Sun~\cite{Miranda:2004nb,Guzzo:2004ue,Friedland:2004pp}, the impact of
NSI at the level of detection has received far less attention and only
qualitative studies have been performed so
far~\cite{Berezhiani:2001rt, Davidson:2003ha}~\footnote{Solar and
reactor neutrino fluxes are unaffected by the class of NSI which
typically arise in models of neutrino mass.}.

Therefore, it seems timely and interesting to investigate in more
detail NSI trying to fill this gap in the literature. Our main aim is
then to perform a quantitative analysis of the impact of NSI in solar
neutrino phenomenology consistently taking into account their impact
both on propagation and on detection processes.  The simultaneous
inclusion of NSI effects in both processes unavoidably renders the
computational analysis very demanding since for each choice of the NSI
couplings, one has to convolve the oscillation probability with
cross-section of the relevant process.  For definiteness in this work
we have restricted our study to the following situation: I) We have
considered only non universal (NU) flavor conserving interactions
neglecting flavor changing neutral current interactions (FCNC).  II)
We have considered interactions only with electrons ($f = e$).  III)
We have performed our analysis switching on the interaction for one
neutrino flavor at a time. IV) We do not consider NSI of
  $\nu_\mu$ with electrons since the current bounds in this
  case~\cite{Barranco:2007ej} ($-0.033\leq\varepsilon^{L}_{\mu\mu}\leq
  0.055$~, $-0.040\leq\varepsilon^{R}_{\mu\mu}\leq 0.053$~) are
  stronger than the attainable sensitivity from our solar analysis.

A final remark is in order. In general, one should also
  consider the possible simultaneous presence of FCNC, and include NSI
  with up and/or down quarks\footnote{Limits on NSI involving up
      and down quarks have already been reported in the
      literature~\cite{Miranda:2004nb,Friedland:2004ah,fornengo:2001pm}.}.
  We have not performed such a general analysis since the number of
  parameters would disproportionally increase. Although considering
  only flavor preserving NSI with electrons may seem somewhat
  reductive, we deem that a model-indpendent detailed study of this
  specific case may provide particular insight and may be useful for
  future, more complete, studies.

The paper is organized as follows.  In Sec.~II we discuss the impact
of NU non-standard interactions on propagation properties providing
quantitative constraints on their amplitude.  In Sec.~III we consider
the effect of NSI on the elastic scattering cross section. In section
IV we discuss the general case in which we simultaneously include NSI
{\it both} in the propagation {\it and} in detection of electron
neutrinos.  In Sec.~V we show analogous results for the case of
$\tau$ neutrinos. Finally, in Sec.~VI we trace our conclusions.

\section{Non-standard propagation} 
\label{Section:prop}

In this section we introduce the basic formalism describing neutrino
propagation in the presence of non-standard interactions and derive
quantitative bounds on the amplitude of the effective non-universal
couplings.  These bounds will be an important ingredient to interpret
the results of our full analysis presented in Sec.~IV and Sec.~V
where we consider the interplay of NSI effects in propagation and
detection processes.

Here and in the following, we assume the standard parametrization for
the lepton mixing matrix~\cite{schechter:1980gr}, within the
convention adopted by the Particle Data Group~\cite{Amsler:2008zz},
setting the small mixing angle $\theta_{13}$ to zero for the sake of
simplicity. For $\theta_{13}=0$, standard oscillations in the
$\nu_e\to\nu_e$ channel probed by long-baseline reactor (KamLAND) and
by solar neutrino experiments are driven by only two parameters: the
mixing angle $\theta_{12}$ and the neutrino squared mass difference
$\Delta m^2_{21}=m^2_2-m^2_1$.  In the flavor basis, the evolution of
neutrinos can be written as
%........................................................................
\begin{equation}\label{2nuevol}
 i\, \frac{d}{dx}\left(\begin{array}{c}\nu_e\\ \nu_a\end{array}\right) = H
 \left(\begin{array}{c}\nu_e\\ \nu_a \end{array}\right)\ ,
\end{equation}
%........................................................................
where $\nu_a$ is a linear superposition of $\nu_\mu$ or $\nu_\tau$, 
and $H$ is the total hamiltonian 
%........................................................................
\begin{equation}
H = H_\mathrm{kin} + H_\mathrm{dyn}^\mathrm{MSW} + H_\mathrm{dyn}^\mathrm{NSI}  
\label{H_tot}
\end{equation}
%........................................................................
split as the sum of the kinetic term, the standard Mikheev-Smirnov-Wolfenstein (MSW) matter
term~\cite{wolfenstein:1978ue,mikheev:1985gs} and of a new,
NSI-induced, matter term~\cite{valle:1987gv}. The kinetic term depends
on the mixing angle $\theta_{12}$, on the squared mass difference
$\Delta m^2_{21}=m^2_2-m^2_1$, and on the energy $E$ \ as
%........................................................................
\begin{equation}
H_\mathrm{kin} = \frac{k}{2}
\left(\begin{array}{lc}
-\cos 2\theta_{12}  ~~ & 
\sin 2\theta_{12} \\
~~~\sin 2\theta_{12} &
\cos 2\theta_{12}
\end{array} \right)
\label{msw-hamiltonian}
\end{equation}
%........................................................................
where $k=\Delta m^2_{21}/2E$ is the neutrino oscillation wave number.
The standard (MSW) interaction term can be expressed as
%........................................................................
\begin{equation}
   H_\mathrm{dyn}^\mathrm{MSW} = V(x)
   \left( \begin{array}{cc}
       1 & 0 \\  0 & 0
   \end{array}\right) \,
   \label{msw-ham}
\end{equation}
%........................................................................
where $V(x) =\sqrt 2 G_F  N_e(x)$  is the effective potential
induced by interaction with the electrons with number
density $N_e(x)$. The NSI term can be cast in the form  
%........................................................................
\begin{equation}
   H_\mathrm{dyn}^\mathrm{NSI} = V(x)
   \left( \begin{array}{cc}
       0 & \varepsilon \\ \varepsilon & \varepsilon'
   \end{array}\right) \,
   \label{nsi-ham}
\end{equation}
%.......................................................................
where  $\varepsilon$ and $\varepsilon'$  are two effective parameters that,
neglecting $\eps_{\alpha\mu}^{f P}$~, 
are related with the vectorial couplings by:
%.......................................................................
\begin{equation}
   \eps = - \sin\theta_{23}\,\eps_{e\tau}^{e V} \qquad
   \epp = \sin^2\theta_{23}\,\eps_{\tau\tau}^{e V} -
   \eps_{ee}^{e V}\,. 
   \label{eps}
\end{equation}
%.......................................................................
In the present work we focus on the flavor conserving non-universal
(NU) couplings, setting the flavor-changing off-diagonal coupling
$\eps = 0$.  Hence, in the treatment of solar neutrino propagation, in
addition to the mass-mixing parameters we include the coupling
$\eps'$.

In our numerical analysis we have included the data from the
radiochemical experiments Homestake~\cite{cleveland:1998nv},
Sage~\cite{abdurashitov:2002nt} and
GALLEX/GNO~\cite{hampel:1998xg,Altmann:2005ix,gallex-nu08}, from Super-KamioKande
(Super-K)~\cite{fukuda:2001nj,fukuda:2001nk,fukuda:2002pe},  from
all three phases of the Sudbury Neutrino Observatory
(SNO)~\cite{ahmad:2001an,ahmad:2002jz,ahmad:2002ka,Ahmed:2003kj,Aharmim:2005gt,Aharmim:2008kc},
and from Borexino~\cite{Arpesella:2008mt}.
We have also included the latest KamLAND data%presented in the Neutrino
%2008 conference
~\cite{:2008ee} using a threshold of $2.6~\mathrm{MeV}$,
which allows us to neglect the contribution of low-energy
geo-neutrinos.

It is worth noticing that although we have incorporated both standard
and non-standard matter effects, due to the low matter density of the
Earth crust, they have only a negligible effect in KamLAND, for the
range of parameters we are considering.  Therefore the inclusion of
KamLAND in the analysis has the important effect of determining the
solar mass-mixing parameter, independently of the non-standard
interaction parameters.

\begin{figure}
\includegraphics[scale=0.60, bb= 100 100 510 720]{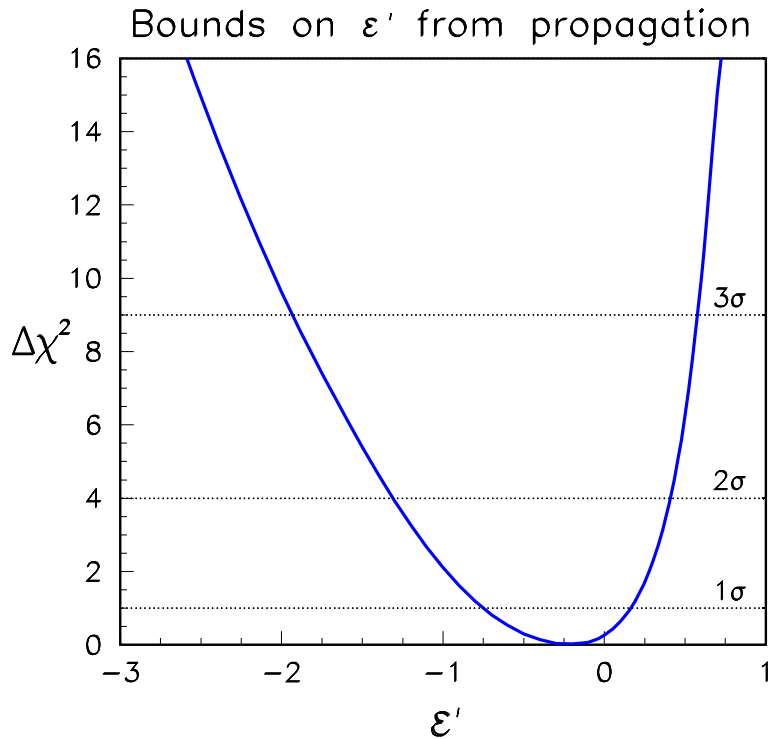}
\vspace*{-2.0cm} \caption{ \label{1pan_prop}  Constraints on the effective 
amplitude characterizing NU non-standard interactions in propagation. }
\end{figure}

In Fig.~\ref{1pan_prop} we show the constraints we obtain on the
parameter $\eps'$ from the solar neutrino data in combination with
KamLAND after marginalization over the two mass-mixing parameters. We
can qualitatively explain these bounds as follows. We notice that,
since the term containing the effective NU coupling is diagonal, it is
formally equivalent to a redefinition of the potential V~\footnote{As
shown in~\cite{Bahcall:2005va}, the uncertainty in the solar
composition leads to a small uncertainty on the electron neutrino
density (and then on the potential V).  In the region relevant for
adiabatic transitions of solar neutrinos $R < 0.6$ (in units of solar
radii) this can be quantified as less than 2\%, hence negligible in
the context of our analysis.},
%.......................................................................
\begin{equation}
   V(x) \to (1 - \eps')V(x)
   \,.
\end{equation}
%.......................................................................
%%%%%%%%%%%%%%%%%%%%%%%%%%%%%
%%%%%%%%%%%%%%%%%%%%%%%%%%%%%%%%%%%%%%%%%%%
In the LMA region the propagation is adiabatic so that, up to small
Earth matter effects, the $\nu_e$ survival probability is given by the
simple formula
%.......................................................................
\begin{equation}
P_{ee}=\frac{1}{2}\left(1+ \cos 2\tilde\theta_{12}(x_0)
\cos 2\theta_{12}\right) \,,
\label{sol_osc}
\end{equation}
%.......................................................................
where $\tilde \theta_{12}(x_0)$ is the energy-dependent
effective mixing angle in matter at the production point $x_0$
(see, e.g., \cite{Smirnov:2007pw} and references therein),
%.......................................................................
\begin{equation}
\cos 2\tilde\theta_{12}(x_0) = 
\frac{\cos 2\theta_{12} - V(x_0)/k}
{\sqrt{\left(\cos 2\theta_{12}  - V(x_0)/k \right)^2 + \sin^2 2 \theta_{12} }}
\,.
\label{cos_mat}
\end{equation}
%.......................................................................
From the equations above we see that the survival probability depends
on the potential $V(x)$ through the ratio $V/k$, and a rescaling of
$V$ can be compensated by a rescaling of the wave number $k$, which
for a fixed neutrino energy implies a rescaling of the value of
$\Delta m^2_{21}$ preferred by data. Therefore, in the presence of a
small NU coupling the LMA solution moves upward ($\eps'<0$) or
downward ($\eps'>0$) in the mass-mixing parameter space (not shown).
Now we note that in the absence of non-standard interactions the value
of $\Delta m^2_{21}$ preferred by solar data is in agreement to the
one identified with high precision by KamLAND. Hence, the presence of
the additional non-standard effects tends to spoil this agreement and
the tension arising between solar and KamLAND effectively constrains the
amplitude of $\eps'$~\footnote{This behavior was indeed already
  noticed in Ref.~\cite{Fogli:2003vj}, where upper bounds on possible
  deviations from the standard amplitude of the MSW interaction
  potential were considered.}.  It is interesting to note that the
constraints on such parameter have now reached the ``sensitivity
limit'' attainable by KamLAND high precision
measurements~\cite{:2008ee}.  Indeed, we have checked that the
constraints that one would obtain fixing the $\Delta m^2_{21}$ at the
best fit obtained by KamLAND are practically equivalent to those we
obtain by exact marginalization.  The freedom for $\eps'$ is
essentially determined by the range of $\Delta m^2_{21}$ allowed by
the solar data {\em alone}.  Indeed, by varying the value of $\eps'$, the
wide solar LMA solution smoothly ``slides'' over the thin $\Delta
m^2_{21}$ region determined by KamLAND.

We observe that while for small deviations around the
standard value ($\eps'=0$) the bounds are symmetrical, for larger
amplitudes the constraints becomes asymmetrical, i.~e. stronger for
positive values of $\eps'$. This behavior is due to the typical shape
of the solar LMA solution (see for
example~\cite{Fogli:2005cq,Schwetz:2008er}) which is more (less)
elongated towards large (low) $\Delta m^2_{21}$ values.  Indeed, the solar
LMA solution is strongly limited from below by the (non) observation
of day-night asymmetry in Super-K and SNO and it is constrained in the
upper part essentially by the CC/NC ratio measured by SNO.  This
asymmetric behavior will be relevant when considering (see Sec.~IV
and~V) the interplay among the limits coming from non-standard
propagation with those coming from non-standard detection.

\section{Non-standard detection}
\label{sec:non-stand-detect}

Non-standard couplings of neutrinos with electrons affect the elastic
scattering $(\nu_a e \to \nu_a e)$ process modifying the number of
events and their spectral distribution expected in the Super-K
detector and to a much lesser extent in the SNO detector. In principle they also affect the Borexino
spectrum but we have checked that the current statistics is (still) 
too low to compete with Super-K. 

The standard differential cross section for $(\nu_a e \to \nu_a e)$
scattering processes has the well known form
%......................................................................
\begin{equation}
\label{eq:cs}
\frac{d\sigma^{\rm std}_{a}}{dT} (E_\nu, T_e)=
{2 G_F^2 m_e \over \pi}
\left[(g_1^a)^2  +  (g_2^a)^2\left(1-\frac{T_e}{E_{\nu}}\right)^2-
g_1^a  g_2^a \frac{m_e T_e}{ E^2_{\nu}}\right],
\end{equation}
%......................................................................
where $m_e$ is the electron mass, $E_\nu$ is the incident neutrino
energy, $T_e$ is the electron recoil energy.  The quantities $g_1^a$
and $g_2^a$ are related to the SM neutral current couplings of the
electron $g_L^e = -1/2 + \sin^2\theta_W$ and $g_R^e = \sin^2\theta_W$,
with $\sin^2\theta_W = 0.23119$~\cite{Amsler:2008zz}\footnote{For our
  numerical analysis, instead of this simple tree level expression, we
  also include the radiative corrections given in
  Ref.~\cite{Bahcall:1995mm}.}.  For $\nu_{\mu,\tau}$ neutrinos, which
take part only in neutral current interactions we have $g_1^{\mu,\tau}
= g_L^e$, $g_2^{\mu,\tau} = g_R^e$, while for electron neutrinos both
charge current (CC) and neutral current (NC) interactions are present
and $g_1^e = 1+ g_L^e$, $g_2^e = g_R^e$. In the presence of NU
non-standard interactions the cross section can be written in the same
form of Eq.~(\ref{eq:cs}) but with $g_{1,2}^a$ replaced by the
effective non-standard couplings $\tilde g_1^a = g_1^a + \eps_{a
  a}^{eL}$ and $\tilde g_2^a = g_2^a + \eps_{a a}^{eR}$.

Strong limits can be placed on $\nu_\mu$ interactions with
electrons~\cite{Barranco:2007ej}
($-0.033\leq\varepsilon^{L}_{\mu\mu}\leq 0.055$~,
$-0.040\leq\varepsilon^{R}_{\mu\mu}\leq 0.053$~). In contrast, the
constraints on the other two NU couplings are rather
loose~\cite{Barranco:2007ej}.
Therefore in our analysis we can safely neglect NSI with muons of
either helicity, and focus in what follows on possible non-standard
couplings of $\nu_e$ and $\nu_\tau$. In addition we have performed our
analysis switching on one flavor non-standard interaction at a time,
due to computational limits. Indeed, already in this simple case we
must consider as additional parameters $ \eps_{a a}^{eL}$ as well as $
\eps_{a a}^{eR}$ at the level of detection, and their sum at the level
of propagation.

Before introducing our numerical results it is worth discussing the
qualitative behavior one expects when NU interactions are present in
the detection process.  We first observe that for the high energy
Boron neutrinos (which are relevant for Super-K) MSW matter effects
dominate and the survival probability is approximately $P_{ee} \sim
\sin^2{\theta_{12}} \sim 1/3$. Furthermore, the transition
probabilities to the other flavors are approximately equal ($P_{e\mu}
\sim P_{e\tau} \sim 1/3$) since the admixture of $\nu_\mu$ and
$\nu_\tau$ neutrinos is determined by the nearly maximal
``atmospheric'' mixing
angle~\cite{Maltoni:2004ei,Fogli:2005cq,Schwetz:2008er,Fogli:2008ig}
($\sin^2{\theta_{23}} \sim 0.5$).  Hence, up to small earth matter
effects, an approximately equal admixture of the three neutrino
flavors arrives at the Super-Kamiokande detector.
Therefore from Eq.~(\ref{eq:cs}) one can expect the following general
features: I) In both cases of $\nu_e$ and $\nu_\tau$ interactions, a
deviation of the L-type coupling should mostly affect the total rate
through the first term in Eq.~(\ref{eq:cs}).  II) The relative
contribution of the first term in the cross section is almost one
order of magnitude larger for $\nu_e$ compared to $\nu_\tau$
($(g_1^e)^2/(g_1^\tau)^2 \simeq 7$). Thus we expect this feature to be
reflected in a reduced sensitivity to $\eps^{eL}_{\tau\tau}$ compared
to $\eps_{ee}^{eL}$.  III) Deviations of the R-type coupling will
instead modify the expected energy spectrum through the second
term and (to a lesser extent) through the third term. 
IV) The value of $g_2^a$ is identical for $\nu_e$ and $\nu_\tau$
and we expect comparable sensitivities for the $\eps_{ee}^{eR}$,
$\eps^{eR}_{\tau\tau}$ effective couplings coming from the Super-K
spectral information. V) The third term (proportional to $g_1^a g_2^a$)
is suppressed by the (energy dependent) factor $m_eT_e/E_\nu^2$,
and should induce non negligible effects only in the case of 
electron neutrinos ($a=e$) since in this case $g_1^a$ is bigger
($g_1^e \sim  0.73$ in the standard case).

\section{Constraints on electron neutrino interactions}
\label{sec:constr-electr-neutr}

In this section we present the numerical results of our analysis in
the presence of NU couplings of $\nu_e$ with electrons. With this aim
we have performed a joint analysis of solar and KamLAND data in the
$(\Delta m^2_{21},\,\sin^2\theta_{12}, \, \eps_{ee}^{eL}, \eps_{ee}^{eR})$
parameter space, taking into account that only the vectorial
combination $\eps_{ee}^{eV} = \eps_{ee}^{eL} + \eps_{ee}^{eR}$ 
is involved in the propagation. Moreover, we have limited our scan in the L-type NSI
parameter, $\eps_{ee}^{eL}$, to the range $(-0.3,0.3)$. Although a
degeneracy in the value of this parameter appears when one includes
only the $\nu e$ scattering data~\cite{Barranco:2005ps}, allowing for
NSI values as large as $\eps_{ee}^{eL}=-1.5$, these values turn
out to be forbidden when one also includes the LEP data, as shown in
Ref.~\cite{Barranco:2007ej}.

\begin{figure}
\includegraphics[width=0.95\textwidth,clip]{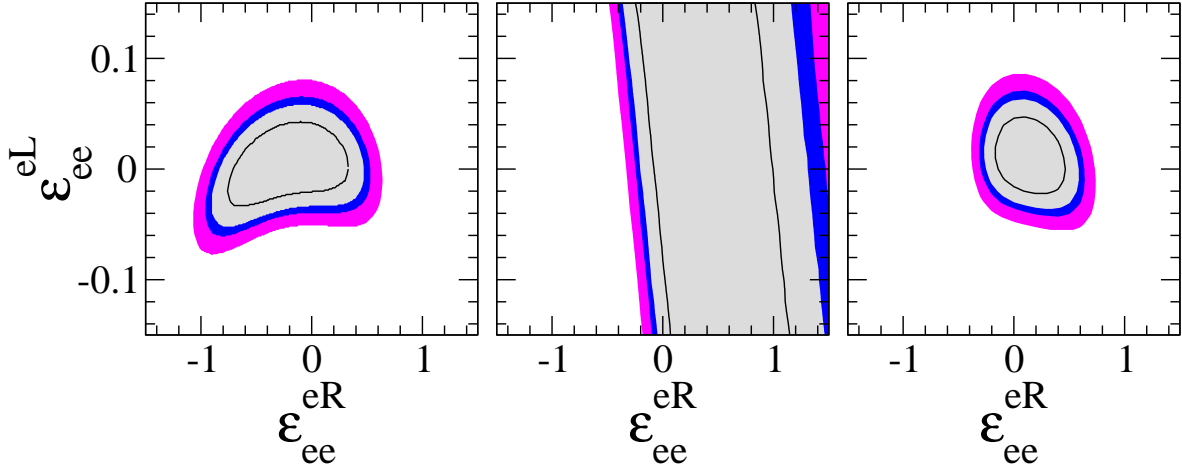}
%\vspace*{-8.0cm} 
\caption{ \label{3pan_ee} Constraints on the electron neutrino
  non-standard interactions. Bounds at 68\%, 90\%, 95\% and 99\% for 2
  d.o.f..  In the left panel non-standard effects are included only in
  the detection, in the middle panel only in propagation and in the
  right panel the effects are included in both processes.}
\end{figure}

In the three panels of Fig.~\ref{3pan_ee} we show the regions allowed
in the plane $[\eps_{ee}^{eL}, \eps_{ee}^{eR}]$ where the mass-mixing parameters
have been marginalized away.  In the left panel we show the region
allowed when we switch on the non-standard effects {\it only} in the
detection process. The sensitivity to deviations of the
L-type coupling is higher than the R-type
sensitivity (notice the different scale used for the two
parameters). This behavior follows from the fact that the most
important effect of $\eps_{ee}^{eL}$ arises from the first term in
Eq.~(\ref{eq:cs}) and approximately consists in an energy independent rescaling of
the cross section. This in turn leads to deviations of the predicted
theoretical values of the total Super-K rate which are rejected by all
the remaining solar data. To better understand this point we note
that, if only Super-K data were included in the analysis, large
deviations of the total cross section could be allowed since they
could be compensated by a rescaling of the theoretical boron flux
which is still uncertain at the $\sim 20\%$ level. However, the
combination of the Super-K data with the other solar neutrino
experiments drastically improves the sensitivity to $\eps_{ee}^{eL}$.  In
particular SNO plays a crucial role in this respect, limiting possible
departures of the total Super-K rate in two ways. First, the NC
measurement provides a direct measurement of the boron flux in
agreement with the SSM prediction to within $\sim 6\%$ or so,
effectively reducing the allowed space for a possible rescaling of the
boron flux. Second, the precision measurement of the SNO CC rate
imposes a further constraint on the Super-K ES rate.
%%%%%%%%%%%%%%%%%%%%%%%%%%%%%
%%%%%%%%%%%%%%%%%%%%%%%%%%%%%%%%%%%%%%%%%%%%

As already observed in the previous section, the constraints on the
R-type coupling come from the spectral information obtained in the
Super-K experiment. Current Super-K data are consistent with the
spectrum predicted for standard cross-section, while still allowing
for appreciable deviations. Therefore the limits on the R-type
coupling are looser compared with those obtained on the L-type one
(note the different scale used for $\eps_{ee}^{eR}$ and
$\eps_{ee}^{eL}$).  We observe that the ``barycenter'' of the
allowed region is slightly shifted toward negative values of
$\eps_{ee}^{eR}$ ($\sim-0.2$). For such values the coefficient $g_2^e
\sim 0$ and both the second and third (energy dependent) terms in
Eq.~(\ref{eq:cs}) tend to vanish indicating a slight preference of the
data for an energy independent cross section. We also observe how the
allowed region is elongated towards negative values of both
non-standard L-type and R-type couplings indicating that in this
region of the parameter space a degeneracy exists between the second and
the third term in Eq.~(\ref{eq:cs}). Indeed, the second term tends to
give a negative tilt to the Super-K energy spectrum which is
counterbalanced by the positive tilt induced by the third one (indeed
its coefficient is positive in this parameter region since $g_2^e$
assumes negative values).

In the middle panel of Fig.~\ref{3pan_ee} we report the constraints
obtained when we include non-standard effects only in neutrino
propagation, as already discussed in Sec.~\ref{Section:prop}. In this
plane these constraints are represented by diagonal bands delimited by
lines corresponding to constant values of the vectorial coupling. This
plot clearly shows how these constraints are different and
complementary to those coming from detection.

In the third panel we show the allowed region obtained by the full
global analysis, where we {\em simultaneously} include non-standard
effects in detection and in propagation.  The effect of including NU
couplings in both processes leads to an appreciable reduction of the
allowed region evidencing a high complementarity and synergy of the
two kinds of constraints, which effectively turns the global allowed
region into a ``round'' shape.

It is interesting to observe that the allowed region in the third
panel looks like just a na\"ive combination of the two regions
determined {\em separately} only by detection and only by propagation.
This result is important since, {\em a priori}, one would in principle
expect a possible degeneracy among non-standard effects induced at the
level of detection and those induced at the level of propagation.  In
particular, some region of the parameter space could exist where
non-standard effects in detection could counterbalance those induced
in the propagation process (and vice versa.)  Our analysis shows, {\em
  a posteriori}, that such a degeneracy is instead absent. One can
qualitatively understand this behavior noting that, although
non-standard propagation effects could in principle partially undo the
modifications induced by the non-standard detection in Super-K, their
presence would unavoidably spoil the agreement of {\em all the other}
experimental results (Cl, Ga, and SNO) with their respective
theoretical predictions (which are all well described by standard
propagation.)

We close this section quoting the range allowed [at $90\%$ C.L. (2 d.o.f.)]
for the amplitude of the non-universal R-type coupling of electron neutrinos
with electrons,
%..............................................................................
\begin{eqnarray} 
-0.27< \eps_{ee}^{eR} < 0.59\,,
\end{eqnarray} 
%..............................................................................
and for the L-type one, 
%..............................................................................
\begin{eqnarray} 
-0.036 < \eps_{ee}^{eL} < 0.063\,.
\end{eqnarray} 
%..............................................................................
We observe that our limits are comparable
with those found by laboratory experiments~\cite{Barranco:2007ej}.

\section{Constraints on tau neutrino interactions}
\label{sec:constr-tau-neutr}

In this section we present the numerical results of the analysis in
the presence of non-universal couplings of $\nu_\tau$ with electrons. As
in the case of the electron neutrinos presented in the previous
section, also in this case we have performed a joint analysis of solar
and KamLAND reactor data in the $(\Delta m^2_{21},\,\sin^2\theta_{12},\,
\eps_{\tau\tau}^{eL}, \eps_{\tau\tau}^{eR})$ parameter space, again
taking into account that only the vectorial combination of the chiral
couplings enters the propagation.  
In contrast to the case considered in the previous section, for the
$\eps_{\tau\tau}^{eL}$ case the analysis is performed for a wider
range than considered for $\eps_{ee}^{eL}$, since the current
laboratory constraints are too weak to resolve the degeneracy pattern~\cite{Barranco:2005ps}.

Note that in the present case the signal observed in the Super-K
experiment is the sum of the standard contribution due to scattering
of the three neutrino flavors, and of an additional nonstandard
contribution due to the interaction of $\tau$ neutrinos with electrons
through the neutral current. These neutrinos originate from solar
neutrino oscillations into a state $\nu_a$ which we approximate as an
equal mixture of $\nu_\mu$ and $\nu_\tau$, corresponding to maximal
"atmospheric" mixing angle and zero $\theta_{13}$.

\begin{figure}
\includegraphics[width=0.95\textwidth,clip]{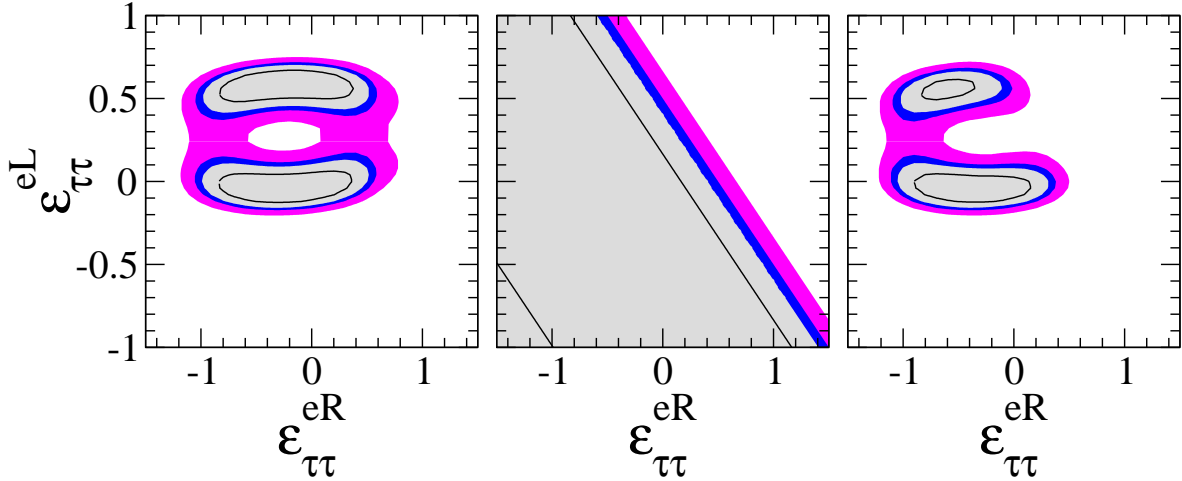}
%\vspace*{-8.0cm} 
\caption{ \label{3pan_tau} Constraints on the $\tau$ neutrino
 non-standard interactions. Bounds at 68\%, 90\% 95\% and 99\% for 2
 d.o.f..  In the left panel non-standard effects are included only in
 the detection, in the middle panel only in propagation and in the
 right panel the effects are included in both processes. Notice the
 different scale for the left coupling with respect to the case of
 electron neutrinos presented in Fig.2. }
\end{figure}

Figure~\ref{3pan_tau} is analogous to Fig. 2 but with the three panels
showing respectively the regions allowed in the $[\eps_{\tau\tau}^{eL},
\eps_{\tau\tau}^{eR}]$ plane. Notice that in this case the scale of the
L-type coupling is different from the case of electron neutrinos, being
almost an order of magnitude larger.
In the first panel, the ``two-island'' behavior is a manifestation of
the degeneracy pattern which exists for the electron
case~\cite{Barranco:2005ps} and which is not fully lifted by our
current global analysis. It is clear from Eq.~(\ref{eq:cs}) that the
neutrino electron cross section is symmetric under the simultaneous
transformation $g^a_1\to -g^a_1$ and $g^a_2\to -g^a_2$.  Moreover, the
last term , already small due to the ratio $m_e/E_\nu$, is further
suppressed compared with the electron neutrino case since its
coefficient $g^\tau_1 g^\tau_2$ is now smaller. Therefore, there is
actually an approximate symmetry under separate changes in the sign of
$g^a_{1,2}$. In our case this can be achieved by setting, for
instance, $\varepsilon_{\tau\tau}^{eL}= -2g^\tau_{1}\simeq0.54$, which
effectively amounts to the transformation
$\tilde{g}^\tau_{1}=g^\tau_{1}+\varepsilon^{eL}_{\tau\tau}\to-g^\tau_{1}$.
As can be seen in Fig.~\ref{3pan_tau}, our global data analysis is
already able to resolve this degeneracy at 99\% C.L., but is not able
to resolve the same degeneracy for the $\varepsilon_{\tau\tau}^{eR}$
case.
%
%%%%%%%%%%%%%%%%%%%%%%%%%%%%%
%%%%%%%%%%%%%%%%%%%%%%%%%%%%%%%%%%%%%%%%%%%
%
As in the case of interaction with electron neutrinos treated in the previous
section we find that the ``barycenter'' of the allowed region is slightly shifted
toward negative values of the L-type parameter, again indicating a weak
preference for a energy independent differential cross section
(see comments in Sec.~\ref{sec:constr-electr-neutr}).

In the middle panel, we show the constraints obtained including non
standard effects only in propagation.  We observe that in this case
(see eq.~\ref{eps}) we have $\eps_{\tau\tau}^{eV} =
\eps'/\sin^2{\theta_{23}} \simeq 2 \eps'$, explaining the reduced
sensitivity to the vectorial coupling.
 Finally, the right panel is obtained, as before, by consistently
 including non standard effects both in neutrino detection as well as
 in propagation. As for the case of electron neutrinos discussed in  
 Sec.~\ref{sec:constr-electr-neutr}, the full analysis clearly shows
 the complementarity among the constraints coming from
 detection and propagation and the absence of any possible
 degeneracy between the two effects. We find the following $90\%$ C.L. (2 d.o.f.) allowed range of the non-standard
 amplitude of R-type coupling:
%..............................................................................
\begin{eqnarray} 
-1.05 < \eps_{\tau\tau}^{eR} < 0.31\,,
\end{eqnarray} 
%..............................................................................
while two disjoint ranges for the L-type coupling are obtained: 
%..............................................................................
\begin{eqnarray} 
-0.16 < \eps_{\tau\tau}^{eL} < 0.11\,, ~~~~~~~~~~~~~~~~~~~
0.41 < \eps_{\tau\tau}^{eL} < 0.66\,.
\end{eqnarray} 
%..............................................................................
corresponding to the "two-island'' region discussed above. 
We observe that also in this case our limits are comparable
to the existing laboratory bounds~\cite{Barranco:2007ej}.

\section{Summary and conclusions}

Motivated by neutrino mass extensions of the standard electroweak
model that imply the existence of neutrino non-standard interactions,
we have considered the constraints on the strength of effective
non-universal (NU) flavor conserving four-fermion operators $\nu_\alpha
\nu_\alpha \bar e e $ with electrons, where $\alpha=e,\tau$, that
can be obtained from solar and reactor (KamLAND) neutrino data.
We have consistently taken into account the effect of non-standard
physics both at the level of neutrino propagation, where they modify
the standard MSW behavior, as well as at the level of detection, where
they affect the cross section of neutrino elastic scattering on
electrons.  

Our analysis allows us to trace the following important conclusions:
I) The constraints on NU couplings obtained by detection
and propagation of solar neutrinos are of comparable sensitivity. 
II) The constraints coming from
the two processes are highly complementary and the general analysis
allows considerable restrictions of the parameter space. 
III) The current data seem powerful enough to remove degeneracies
possibly arising among NU couplings at the level of detection
and propagation respectively. 
IV) The limits we find are comparable with those found by means of
other model-dependent searches.

\acknowledgments

This work was supported by Spanish grants FPA2008-00319/FPA and
FPA2008-01935-E/FPA and ILIAS/N6 Contract
RII3-CT-2004-506222. M.A.T. is supported by the DFG (Germany) under
grant SFB-676. A.B. and O.G.M. were supported by Conacyt and the HELEN 
program.

\bibliographystyle{apsrev}
\def\baselinestretch{1}
\bibliographystyle{h-physrev4}
%\bibliography{valle-ref}

\end{document}